# Frictional heating processes and energy budget during laboratory earthquakes

J. Aubry[1*], F. X. Passelègue[2], D. Deldicque[1], F. Girault[3], S. Marty[1], A. Lahfid[4], H. S. Bhat[1], J. Escartin[3] and A. Schubnel[1]

[1]Laboratoire de Géologie, École Normale Supérieure/CNRS UMR 8538, PSL Research University, 24 rue Lhomond, F-75005 Paris, France.

[2]École Polytechnique Fédérale de Lausanne, CH-1015 Lausanne, Switzerland.

[3]Institut de Physique du Globe de Paris, Sorbonne Paris Cité, Université Paris Diderot, CNRS UMR 7154, 1 rue Jussieu, F-75005 Paris, France.

[4]Bureau de Recherches Géologiques et Minières, 3 avenue Claude-Guillemin, BP 36008, F-45000 Orléans, France.

*Corresponding author. Email address: jerome.aubry@ens.fr

# ENERGY BUDGET OF LABORATORY EARTHQUAKES

**Abstract**

During an earthquake, part of the released elastic strain energy is dissipated within the slip zone by frictional and fracturing processes, the rest being radiated away via elastic waves. Frictional heating thus plays a crucial role in the energy budget of earthquakes, but, to date, it cannot be resolved by seismological data. Here we investigate the dynamics of laboratory earthquakes by measuring frictional heat dissipated during the propagation of shear instabilities at typical seismogenic depth stress conditions. We perform, for the first time, the full energy budget of earthquake rupture and demonstrate that increasing the radiation efficiency, i.e. the ratio of energy radiated away via elastic waves compared to that dissipated locally, increases with increasing thermal – frictional – weakening. Using an in-situ carbon thermometer, we map frictional heating temperature heterogeneities – 'heat' asperities – on the fault surface. Combining our microstructural, temperature and mechanical observations, we show that an increase in fault strength corresponds to a transition from a weak fault with multiple strong asperities, but little overall radiation, to a highly radiative fault, which behaves as a single strong asperity.

# ENERGY BUDGET OF LABORATORY EARTHQUAKES

During an earthquake, the elastic strain energy accumulated during the inter-seismic period is released by rapid fault sliding. The resulting ground shaking, which may cause casualties and material damage, depends in large part on the dynamics[1] and the energy budget of rupture propagation[2,3]: $W=E_R+G_c+H$, where $W$ is the total potential (elastic strain) energy and $E_R$ is the energy radiated away by elastic waves. Fracture energy $G_c$ is the energy required to weaken the fault, allowing the fracture to further propagate, and includes all the energy spent to create a newly fresh fault surface[4], form gouge and micro-cracks at the crack tip or near the rupture front[5]. $H$ is the energy dissipated locally by frictional heating[3]. In the last decade, several studies have proposed that frictional heating activates thermal mechanical processes explaining the dynamic evolution of fault strength, such as grain-size reduction and superplastic flow[6,7], thermal pressurization of pore fluid[8,9], flash heating[9,10], frictional melting[11,12] and mineral phase transformations[13,14]. All these mechanisms result in a significant fault weakening during seismic slip[15], through a process that is largely controlled by the temperature evolution on the fault plane, hence raising concern on the current dichotomy between fracture energy and frictional energy.

On the other hand, seismological observations have shown a high variability of rupture speed[2,5,16], stress drop[17,18] and radiation efficiency[2,19], which solely depends on the radiated and fracture energies[12]. However, while the radiated energy is well-estimated using accelerograms[20], fracture energy, inferred from scaling laws and stress drops, has large uncertainties[21,22,23], and frictional heat remains so far 'invisible' to seismological data. In consequence, field studies along active or fossil faults zones[11,24,25,26,27], numerical modeling[28,29,30,31] and laboratory experiments[32,33,34] have been carried out to further constrain the energy budget of earthquakes. Among these approaches, reproducing earthquakes in the laboratory under controlled thermodynamic conditions, as analogues of natural earthquakes[35], may well provide a unique opportunity to understand the physics of the earthquake source and the seismic energy budget. In this paper, we investigate the seismic energy budget during stick-slip events ('laboratory earthquakes') on Westerly granite under confining pressures typical of the brittle crust. Temperature is measured near the fault surface to quantify frictional heat produced during sliding. Amorphous carbon, combined with Raman micro-spectrometry, is used as an in-situ thermometer allowing for the first time to image small-scale temperature heterogeneities on the fault plane.

## Results

**Total heat produced during laboratory earthquakes** Three series of stick-slip experiments (Fig. 1a) were conducted on saw-cut samples of Westerly granite at confining pressures of 45, 90 and 180 MPa, corresponding to crustal depths of about 1.7, 3.4 and 6.8 km, and at a strain rate of $10^{-6}\,s^{-1}$ (see Methods). For each stick-slip, we measured the static stress drop (Supp. Fig. S1a), the co-seismic displacement (corrected for elasticity; Supp. Fig. S1b), and the temperature evolution 5.5 mm away from the fault surface (Supp. Fig. S1c). Within a series of stick-slip events, peak fault friction, static stress drops and co-seismic displacements increase with time. Static stress drops range from 2 MPa to 74 MPa, resulting in co-seismic displacements ranging from 40 μm to 1.2 mm. As expected[34], the stress drop increases linearly with displacement and the relation follows the apparatus stiffness (Fig. 1a). Stick-slip events with large stress drop and displacement, are followed by clear temperature rises ranging from 0.04°C to 0.4°C (Fig. 1b-d).

Temperature measurements are inverted using a simple heat diffusion model to quantify the heat produced co-seismically by frictional sliding (see Methods). The main model assumptions are first a constant co-seismic heat source lasting 20 μs, measured previously as the characteristic time of weakening in these experiments[32]. Second, we assume that the heat produced is dissipated over a layer twice the size of the thermal diffusion length. For the sake of simplicity, we neglect latent heat effects. The main output of the inversion is the amount of heat produced co-seismically, $Q_{th}$, in order to fit the experimental temperature curves shown in Fig. 1b-d (see Methods). The inversion fits relatively well the temperature evolution for experiments conducted at confining pressures of 45 and 90 MPa (Fig. 1b-c). At confining pressure of 180 MPa (Fig. 1d), the model cannot reproduce the data, which may be due to a thermal anomaly longer than the assumed 20 μs, and to latent heat effects not considered in this study. However, the integration of both experimental and modelled curves within the time interval, i.e. a proxy for the heat energy, remains in good agreement even for large stick-slip events.

The total heat produced during stick-slip increases with increasing co-seismic slip (Fig. 2a) and reaches up to 37 kJ/m$^2$ for the largest stick-slip event (Fig. 1d). The comparison with previous measurements of fracture energy performed with the same set-up under similar stress conditions[32] reveals that frictional heat dissipation is systematically larger than the measured fracture energy, and that this difference is largest under low normal and shear stresses. This is expected because under low normal and shear stress conditions, the frictional drop is low, resulting in small stress drop and low fracture energy. However, under higher normal and shear stress conditions, fracture energy and heat become comparable, which is consistent with the low dynamic friction and almost complete stress drops observed at higher normal stress and for larger slips[15,32]. In summary, we confirm here, with two independent measurements of heat and fracture energy, that the large fracture energies measured during stick-slip events at high stress and large slips are dissipated by frictional heat rather than surface creation, because under these high stress conditions, most of the weakening is driven by thermal processes rather than fractures[15].

**Surface melting and heating efficiency** The evolution of fault strength during the initial stage of slip plays a crucial role on the fault weakening processes[36]. Indeed, flash melting occurs at the onset of sliding[9, 37] for a small amount of slip[38] and at highly stressed frictional asperities during rapid slip[9]. As observed in previous studies, the shear resistance resulting from stick-slip motion is controlled by the thickness of the melt layer produced co-seismically[39]. Post-mortem Scanning Electron Microscope (SEM) micrographs indeed reveal the presence of melt on the fault surface (Fig. 3, Supp. Figs. 2 and 3). At 45 MPa, most biotite crystals and only a few plagioclases and potassic feldspars have melted. By contrast, all minerals including biotite, plagioclase, potassic feldspars and even quartz fragments (minimum temperature of fusion of 1650°C) have melted during experiments at 90 and 180 MPa, indicating that the temperature rise on the fault surface was larger during these experiments (Supp. Fig. 2), thus facilitating a complete shear heating (i.e. efficient melt lubrication, Fig. 3) of the fault[11].

The heating efficiency of our laboratory earthquakes shows a clear relation with the co-seismic displacement. As stresses and displacements were measured continuously, the total (elastic strain) energy, $W=(\tau_0+\tau_r)d/2$, where $\tau_0$ and $\tau_r$ are the initial and final shear stress, respectively, and $d$ is the co-seismic displacement, can also be calculated for each individual stick-slip event. The heating efficiency, $R$, defined by the ratio frictional heat produced $Q_{th}$ to total work $W$, decreases with increasing slip (Fig. 2b), which is compatible with a low dynamic friction coefficient during large laboratory earthquakes (Table 1). Conversely, we argue that the true radiation efficiency in our experiments, i.e. $E_R/W$, must increase with increasing slip to values close to 0.9. This is indeed supported by the fact that fracture energy is either negligible compared with heat at low normal stress and/or for small slip, or dissipated by frictional heat at larger stress and/or for larger slip. Our observations are also consistent with other laboratory experiments reporting close to complete stress drops[32, 40] and high (super-shear) rupture speeds[41].

The apparent scaling of the heating efficiency $R$ with the inverse of the square root of slip $d$ ($R \propto d^{-1/2}$; Fig. 2b) has strong implications for seismic efficiency. The heat produced per unit surface scales as $Q_{th} \propto \rho C \Delta T (\kappa \tau)^{1/2}$, where $\rho$ is the bulk density (kg/m$^3$), $C$ the specific heat capacity (J/kg/K), $\Delta T$ the temperature rise on the fault plane (K), $\kappa$ the thermal diffusivity (m$^2$/s) and $t$ the duration of sliding (s), $(\kappa \tau)^{1/2}$ being the thermal diffusion length (m). $W$ scales with the co-seismic displacement, so that $R$ scales as $d^{-1/2}$ only when $Q_{th}$ scales as $d^{1/2}$, i.e. when the slip is linear with $t$. The latter observation implies that, at the scale of our experiments, for similar apparent stress and stress drops, the heating efficiency decreases, and conversely the true radiation efficiency increases, with $d^{-1/2}$. Our experiments thus show that rupture processes become more and more efficient as sliding increases.

Heating efficiency also decreases with cumulative stick-slips events (Fig. 2b), suggesting that less heat is generated by frictional sliding after cumulative displacements. We studied the effects of the number of stick-slips on the fault surface at confining pressure of 90 MPa (Supp. Fig. S4). After a single stick-slip, we document the presence of gouge particles originating from minerals plucked along the fault interface. This gouge is generally composed of feldspars and biotite with some micro-grains of quartz (Supp. Fig. S4). Agglomerated micro- and nano- particles of gouge form strong asperities (exceeding hundreds of microns in length), which did not melt. After several stick-slips, molten gouge particles are found in micro-scale asperities (Supp. Figs. S2 and S4). These flattened melt patch asperities can be reworked by flash melting, which decreases the heating efficiency. Finally, the degree of shear deformation of the gouge and the amount of melt on the fault surface increase together with increasing confining pressure and slip (Supp. Figs. S2 and S4). The effects of cumulative stick-slips (Supp. Fig. S4) show that a given rupture event is controlled by the state of the fault surface, itself affected by the last rupture event through flash or frictional melting of minerals and the reworking of asperities. This, suggest a time-dependent 'memory effect' of the fault surface.

**Imaging the fault surface temperature during slip** To quantify the heterogeneity of frictional heating processes and identify asperities on the fault surface, we developed a carbon thermometer. A specific temperature calibration was performed (Supp. Fig. S5), relying on the progressive carbonization and graphitization of amorphous carbon upon heating under controlled deoxygenated atmosphere, and characterized with Raman Micro-Spectrometry (RMS)[42,43] (Methods). Post-mortem RMS mapping of the amorphous carbon deposits on the fault plane provides fault surface temperature maps for each experiment (Fig. 4). These estimated temperatures are a good proxy for the carbonization kinetics (Methods); the uncertainty on these temperature estimates is difficult to assess, but likely decreases with increasing temperature due to kinetics (Supp. Fig. S5). Furthermore, this technique is unique as it captures the maximum temperature reached after the cumulative effects of several stick-slip events, and it is insensitive to temperatures lower than 700°C. The use of this in-situ thermometer provides several new insights on weakening processes of faulting. First, for all experiments, the extremely high temperature maxima achieved on the fault surface, above 1500°C (Fig. 4), concurs with the SEM micrographs of melt and the degree of shear melting for the three different confining pressures (Supp. Fig. S2). Second, for experiments conducted at confining pressures of 45 and 90 MPa, the high temperature patches are laterally spread along the slip direction, approximately 100 μm wide and can be several hundreds of microns long (Fig. 4). This geometry suggests a genetic link to asperities and striations on the fault plane. Indeed, 'heat asperities' are longer than the maximum amount of co-seismic displacement of any single stick-slip event during these two experiments. They must therefore have remained active during several stick-slip events. Finally, the average bulk temperature attained on the fault plane increases with increasing confining pressure, and exceedes the melting temperature of all minerals (>1500°C) on the fault plane at confining pressure of 180 MPa. In this case, asperities can no longer be recognised, but striations along the sliding direction remain visible.

The mechanical behaviour of asperities and the amount of melt along the fault relate directly to the heat generated during frictional sliding. The percentage of melting on the fault surface can be inferred from both the RMS temperature mapping (Fig. 4) and the thermocouple inversion (Fig. 1b-d). Comparing these two estimates (Supp. Fig. S6) confirms that RMS-mapped asperities must have been reworked during several stick-slip events. This slip-dependent aggradation of asperities over their lifespan is critical in natural crustal faulting as it may lead to more efficient seismic ruptures, stronger ground shacking and likely devastating earthquakes.

When the fault contact area is maximum and the fault plane entirely molten, heating efficiency decreases with increasing co-seismic slip and cumulative displacement (i.e. number of events). With increasing confining pressure and normal stress, we observe the gradual transition from flash melting at asperity contacts to bulk melting over the fault surface. Our laboratory experiments indeed highlight the transition from the mechanical behaviour of a group of asperities, at low normal stress, to that of a single asperity, at high normal stress. This important transition also corresponds to the transition from a weak fault, displaying few contact points, low driving shear stress, low stress drop and high heating efficiency, to a strong fault having high contact area, high driving shear stress, high stress drop, and high radiation efficiency (Fig. 5).

## Conclusions

Our study has evidenced and imaged experimentally how the transition from the dynamics of a group of micro-scale asperities to that of a single asperity influences radiation efficiency. Seismological studies generally infer seismic radiation efficiency, $\eta = E_R/(E_R+G_c)$, ranging from zero for slow earthquakes[19] to one for large earthquakes[44]. Most earthquakes, and in particular along subduction zones, typically exhibit values of $\eta$ between 0.3 and 0.8[44,45]. Our observations suggest that moderate radiation efficiencies arise from rupture propagating along fault planes where only a small portion of the rupture area is stressed (Fig. 5). This stressed rupture area corresponds to asperities, which however contribute largely to the overall radiation efficiency of rupture, while the vast majority of the fault plane, relatively less stressed, shows limited stress drop and radiates poorly. Few earthquakes have been associated with radiation efficiency above one[44], among them the 1992 Landers and 1994 Northridge earthquakes. Slip inversion studies[46,47] have suggested that these earthquakes were dominated by the behaviour of a single asperity whose size is comparable to the fault plane. We argue, on the basis of our experimental data, that these cases correspond instead to earthquakes with an apparent static stress drop much smaller than the dynamic stress drop on the asperity[32] (Fig. 5), as quenched frictional melt may weld the fault and lead to fast fault re-strengthening and strength recovery[48,49].

Our study therefore demonstrates that seismological radiation efficiencies larger than one are possible without the need to invoke frictional heating as a dominant energy sink. On the contrary, under in-situ pressure and temperature conditions, the difference between the complete radiative budget of earthquake rupture (including heat) and the seismological one (neglecting heat) becomes negligible as radiation efficiency increases, at least at the scale of asperities in the laboratory.

## Methods

**Sample characterization and preparation** The rock material is fine-grained Westerly granite, a reference for rock mechanics experiments under crustal conditions[32]. It is frequently found within dykes or sills cutting the gneiss in Rhode Island, USA and is composed of 33% of potassic feldspars, 33% of plagioclase, 28% of quartz, 5% of phyllosilicates (biotite and muscovite) and 1% of accessory minerals (magnetite and sphene)[32]. The mean grain size is 0.75 mm with a total porosity <1%. In the experiments, we used cylindrical samples 80 mm in length and 40 mm in diameter that were sawed to create a weak interface at 30° with respect to the vertical axis (Fig. 1a). First, the contact surface between the two saw-cut samples was polished with a surface grinder to obtain a flat fault plane, and then roughened with grit paper of #240. Second, a carbon layer of rectangular shape (10 mm × 20 mm) with a thickness of few tens of nanometres (<30 nm) was deposited at the centre of the top surface. Third, the thermocouple was inserted in the footwall sample at 5.5 mm away from the contact surface. Finally, the sample was isolated from the confining oil by a neoprene jacket (125 mm long, 5 mm wall thickness).

**Experimental setup** The triaxial oil-loading cell used for the experiments is a an auto-compensated triaxial cell installed at the Laboratoire de Géologie of École Normale Supérieure (Paris, France), which can reach 300 MPa confining pressure[14]. The displacement of the piston is measured by a linear variable differential transducer (LVDT) placed on top of the piston. Axial load is measured via an internal load cell. Maximum load is 717 MPa for 40 mm diameter samples and minimum axial loading rate is 0.01 MPa.s$^{-1}$. Parameters recorded during the experiment include displacement (measured externally, ±0.1 μm), stresses (±0.01 MPa), temperature (±0.01°C) and strains (±10$^{-6}$) at a sampling rate of 100 Hz.

**Temperature estimated by the carbon deposition technique** To image frictional heating processes during laboratory earthquakes, we designed a specific carbon calibration shown in Supp. Fig. S5. We used a carbon coated platinum crucible heated between 700 and 1400°C in an oven under argon flow during 5s. After heating, each carbon deposits was characterised by Raman Micro-Spectrometry (RMS). Carbon spectra, recorded in the 850–2000 cm$^{-1}$ range, were obtained with a Raman micro-spectrometer (Renishaw Invia, UK; argon laser beam of 514.5 nm; LEICA objective with 50x magnification) located at Bureau de Recherches Géologiques et Minières (Orléans, France). Spectra are composed of two principal bands around 1350 and 1600 cm$^{-1}$, named defect band (D band) and graphite band (G band), respectively. The increase of temperature induces an increase of the height of the D band on Raman spectra. The variation of the ratio of heights of D band to H band, $H_d/H_g$, allows identifying carbonization stages and estimating the maximum attained temperature[43]. We observed that the higher the temperature achieved, the higher the ratio. RMS carbon mapping (dimensions of 2 mm × 0.5 mm) was performed post-mortem on top-wall fault specimen after each experiment. Each pixel on the RMS map corresponds to one measured $H_d/H_g$ ratio. Additionally, we used a scanning electron microscope (Zeiss ∑igma, Ecole Normale Supérieure, Laboratoire de Géologie, Paris, France) to control if the area is well melted.

**Temperature measured by the thermocouple** When a stick-slip occurs, the thermocouple (K-type, OMEGA), installed at 5.5 mm away from the fault surface, records the evolution of the temperature at 100 Hz. Temperature data was then used to estimate the energy dissipated into heat during a given stick-slip event using the heat equation, assuming a constant heat source during slip duration ($t_w$=20 μs ). Indeed, the thermocouple gives access to the thermal diffusion at a larger timescale than the duration of the stick-slip event and minimization between the model and the experimental data (Fig. 1b-d) was performed using a least square method, considering a diffusion duration of 20 s (Fig. 1b-d). Parameters used for the inversions are: $\rho$

the density of Westerly granite (2650 kg/m$^3$); $C$ the specific heat capacity of the rock (900 J.kg$^{-1}$.K$^{-1}$); $t_w$ the duration of the event[32] fixed at 20 µs; $w$ the width of the sheared layer (m), defined as $w=(\pi K t_w)^{1/2}$, where $K$ is the thermal diffusivity of Westerly granite (1.10$^{-6}$ m$^2$s$^{-1}$).

## Acknowledgements

The authors would like to thank Y. Pinquier (ENS, Paris, France) and K. Oubellouch (ENS, Paris, France) for their valauble help in laboratory experiments and sample preparation. This manuscript has been greatly improved thanks to the comments and suggestions of S. Bernard (IMPMC, Paris, France) and H. Lyon-Caen (ENS, Paris, France). This study was funded by the European Research Council grant REALISM led by A. Schubnel.

## Author contributions

A.S., H.B., F.G., F.X.P. and J.E conceived the study. J.A, S.M. and A.S. performed the experiments. J.A., F.X.P., H.B. and A.S. performed the temperature inversion and energy budget discussion. J.A., A.L. and D.D. performed the Raman carbon temperature calibration and microstructural analyses. All co-authors participated to the redaction of the manuscript.

## Competing Interests

The authors declare that they have no competing financial interests.

## Materials & Correspondence

Correspondence and requests for materials should be addressed to jerome.aubry@ens.fr.

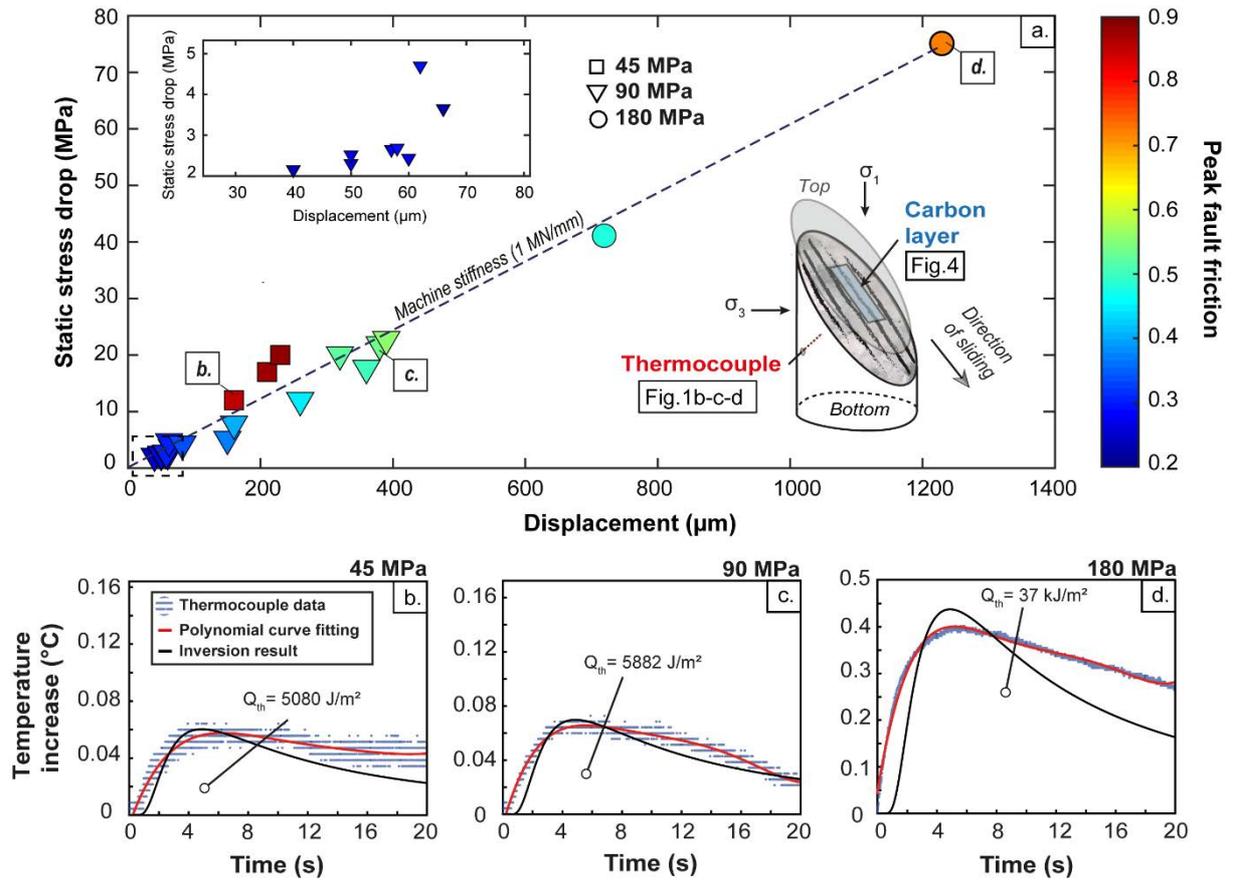

**Figure 1. Mechanical and thermocouple data of laboratory earthquakes**. a) Static stress drop as a function of displacement and peak fault friction for the three confining pressures tested during triaxial experiments (45, 90 and 180 MPa). A thermocouple and an amorphous carbon patch (bottom right-hand corner) are used to measure on the fault surface average temperatures (Fig. 1b-d) and local temperatures maxima (Fig. 4), respectively. The slope of the linear regression of the data points gives the machine stiffness of the triaxial loading cell. b-d) Temperature increase as a function of time for a single stick-slip event at three confining pressures (in narrow bold italic): b) 45 MPa, c) 90 MPa and d) 180 MPa. We invert the thermocouple data to calculate the frictional heat produced on the fault (Table 1, Methods). The amount of heat generated increases with the confining pressure. In blue, raw thermocouple data; in red, polynomial curve fitting; in black, the result of the inversion.

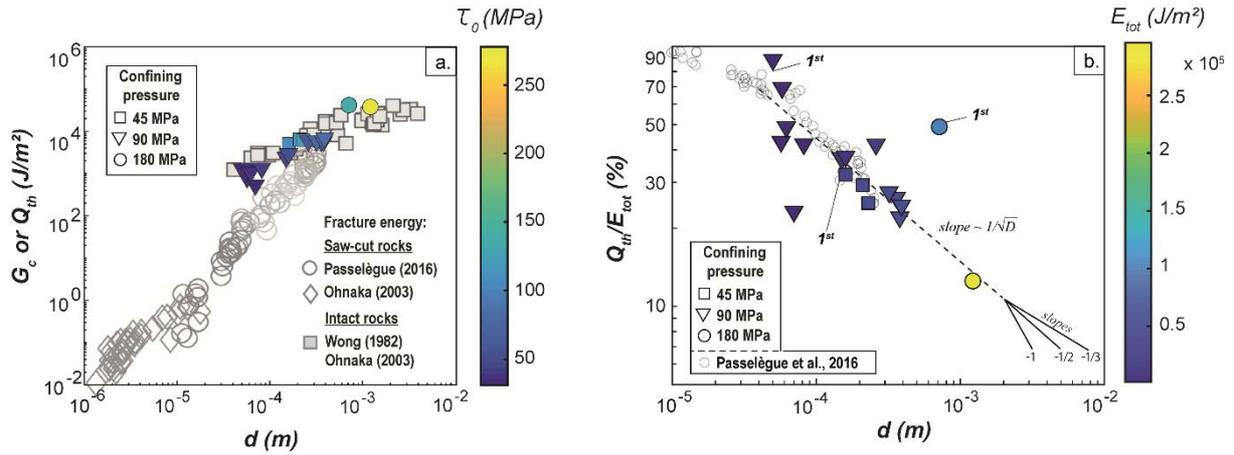

**Figure 2. Frictional heat produced, fracture energy and heating efficiency of laboratory earthquakes.** a) Frictional heat or fracture energy produced during stick-slip as a function of slip and peak shear stress. Experimental values of fracture energy versus slip from the literature[23,27,32] are also plotted for saw-cut rocks[27,32] and intact rocks[23,27]. The total heat produced during stick-slip, $Q_{th}$, increases with increasing co-seismic slip, $d$. b) Heating efficiency as a function of slip and total mechanical work. "1st" indicates the first stick-slip of a series of events, for each experiment. The theoretically expected evolution of heat generation appears to be consistent with our experimental observations. Laboratory earthquakes are less dissipative under large slip and heating efficiency is higher for the first stick-slip of a series of events.

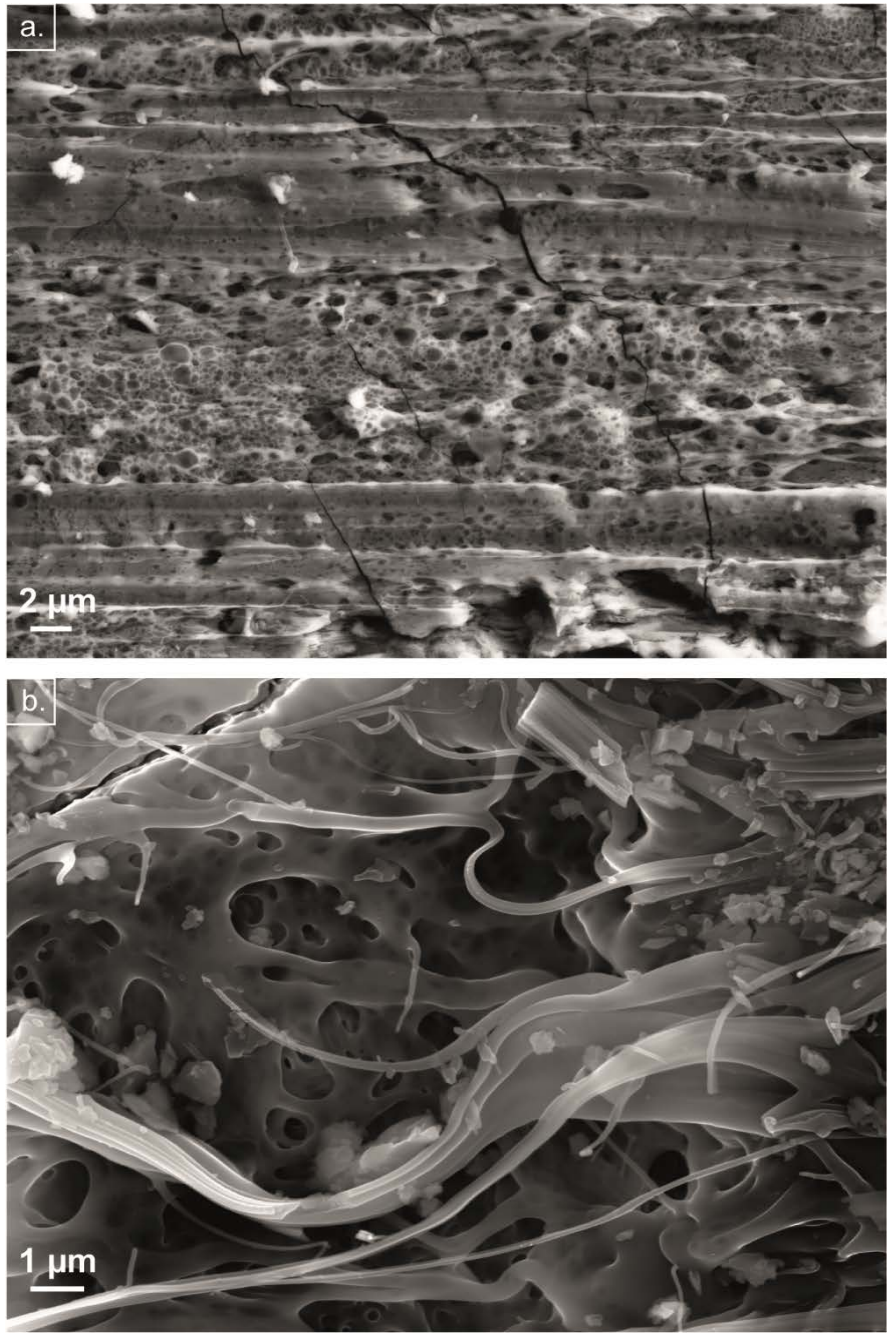

**Figure 3. SEM micrographs of post-mortem surfaces at confining pressure 180 MPa.** At high confining pressure, the fault is entirely molten and true radiation efficiency is very high (around 90%). Striations mapped with the help of RMS mapping in Fig. 4 are shown in a different way here (Fig. 3a). In the molten area, melt distinguishes itself with filamentaous and holey aspect structures (Fig. 3b).

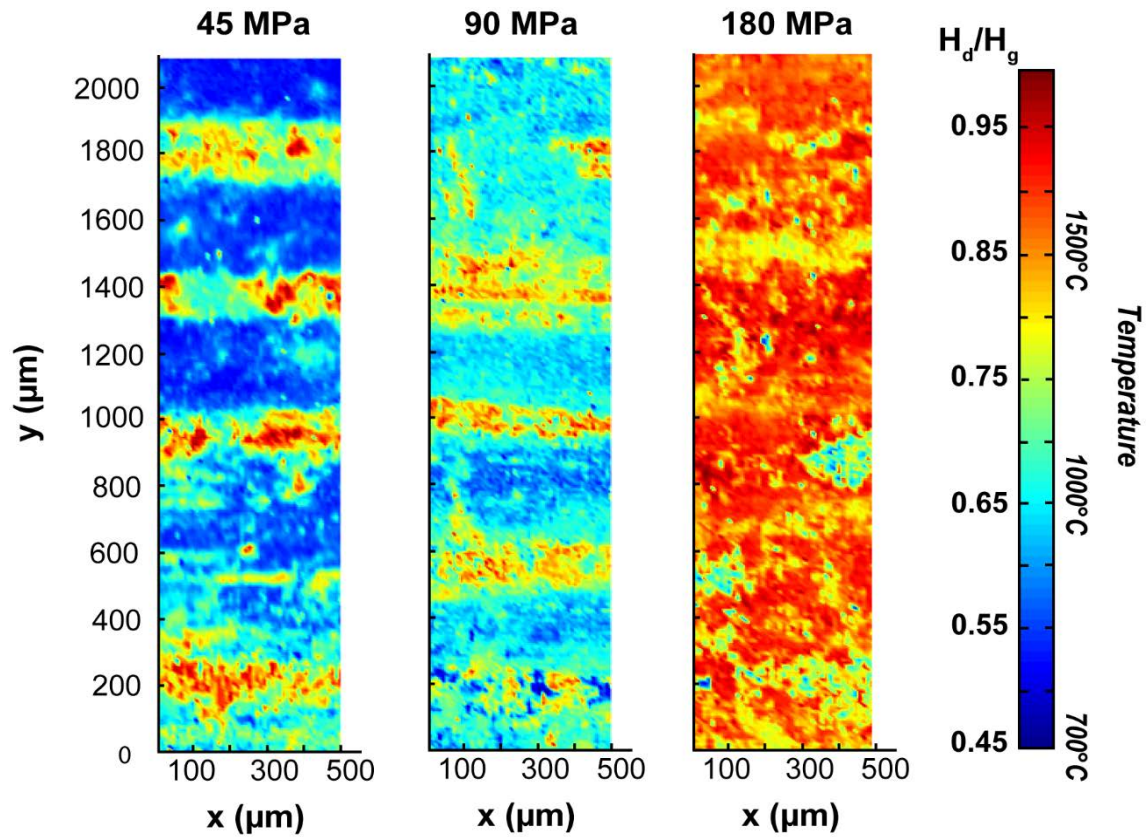

**Figure 4. Temperature maps during frictional sliding for three different confining pressures.** Confining pressures of 45, 90 and 180 MPa are tested. The temperature reached on the fault increases with the confining pressure. During a laboratory earthquake, and over a few microns of slip, asperities reach a heterogeneous and large range of melting temperatures. Each pixel on the temperature maps corresponds to the height ratio of the D and G bands of carbon, $H_d/H_g$, thus allowing to estimate the rise of the temperature[43]. See Methods and Supp. Mat. for details.

ENERGY BUDGET OF LABORATORY EARTHQUAKES

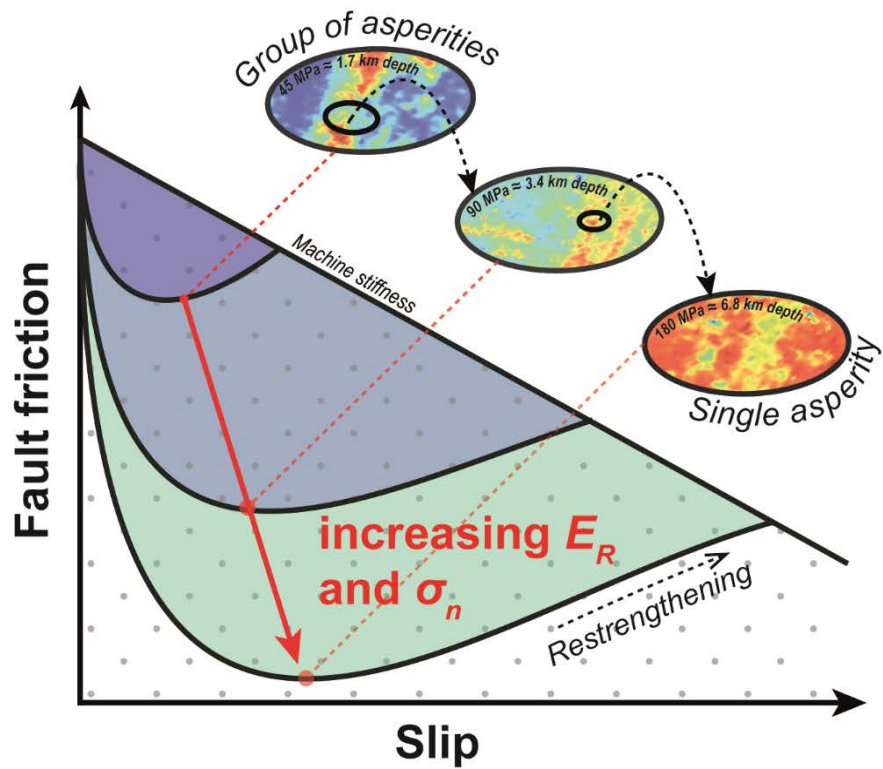

**Figure 5. Schematic of friction evolution by combining thermocouple temperature measurements, carbon temperature estimates, and energy budget.** With increasing confining pressure, the mechanical behaviour evolves from that of a group of asperities to that of a single asperity. When the fault surface is entirely molten, heat generation by frictional processes, and fracture energy, becomes low and radiated energy increases. Coloured regions correspond to the radiated energy, $E_R$.

**Table**

| Confining pressure (MPa) | Number of stick-slip | Normal stress, $\sigma_n$ (MPa) | Peak shear stress, $\tau_0$ (MPa) | Residual shear stress, $\tau_r$ (MPa) | Static stress drop (MPa) | Co-seismic slip, $d$ (mm) | Peak thermocouple temperature increase (°C) | Total mechanical energy, $E_{tot}$ (J/m²)* | Frictional heat produced, $Q_{th}$ (J/m²)* | Heating efficiency, $R$ (%)* |
|---|---|---|---|---|---|---|---|---|---|---|
| 45 | 1 | 104 | 105 | 93,0 | 12,0 | 0,160 | 0,076 | 15840 | 5080 | 32 |
| 45 | 2 | 110 | 113 | 96,0 | 17,0 | 0,210 | 0,103 | 21945 | 6417 | 29 |
| 45 | 3 | 112 | 117 | 97,0 | 20,0 | 0,230 | 0,106 | 24610 | 6149 | 25 |
| 90 | 6 | 107 | 31 | 29,1 | 2,0 | 0,050 | 0,026 | 1505 | 1337 | 89 |
| 90 | 7 | 109 | 33 | 31,8 | 1,7 | 0,057 | 0,021 | 1864 | 802 | 43 |
| 90 | 8 | 111 | 33 | 33,1 | 0,5 | 0,058 | 0,025 | 1932 | 1337 | 69 |
| 90 | 9 | 110 | 34 | 31,8 | 2,4 | 0,070 | 0,022 | 2308 | 535 | 23 |
| 90 | 10 | 111 | 36 | 33,8 | 3,0 | 0,062 | 0,021 | 2186 | 1069 | 49 |
| 90 | 11 | 113 | 40 | 36,8 | 4,0 | 0,082 | 0,026 | 3179 | 1337 | 42 |
| 90 | 12 | 116 | 46 | 40,5 | 5,6 | 0,150 | 0,038 | 6493 | 2406 | 37 |
| 90 | 13 | 121 | 52 | 44,7 | 8,2 | 0,160 | 0,038 | 7800 | 2941 | 38 |
| 90 | 14 | 125 | 59 | 48,3 | 11,3 | 0,260 | 0,073 | 14031 | 5882 | 42 |
| 90 | 15 | 131 | 70 | 53,8 | 16,9 | 0,360 | 0,072 | 22408 | 5882 | 26 |
| 90 | 16 | 134 | 76 | 56,7 | 20,2 | 0,320 | 0,073 | 21368 | 5882 | 28 |
| 90 | 17 | 137 | 80 | 59,8 | 21,0 | 0,380 | 0,077 | 26697 | 5882 | 22 |
| 90 | 18 | 139 | 84 | 61,3 | 23,2 | 0,390 | 0,086 | 28423 | 6951 | 25 |
| 180 | 1 | 264 | 136 | 95,0 | 41,0 | 0,720 | 0,440 | 83160 | 40638 | 49 |
| 180 | 2 | 300 | 278 | 203,0 | 75,0 | 1,230 | 0,401 | 295815 | 36895 | 13 |

**Table 1.** Overview of the mechanical and thermocouple (raw measurements and parameters calculated with the thermocouple inversion model* (note that the distance fault/thermocouple is 5.5 mm). Only data for stick-slips events with an exploitable thermocouple increase signal (i.e. why we begin to stick-slip number 6 in the 90 MPa confining pressure series) are referenced here and used in the study.

# Supplementary Information

**Frictional heating processes and energy budget during laboratory earthquakes**

**Aubry et al.**

The present work is dedicated to the study of frictional heating processes and energy budget during fault sliding. In the following paragraphs, first, we detail the mechanical and thermocouple data of our experiments. Second, we characterise the microstructures of post-mortem faults and discuss the relationship between flash-heating (i.e. flash melting) and fault weakening. Third, we detail the calibration of the carbon technique for temperature estimation, and how comparing carbon and thermocouple methods shed light on the impact of sliding (i.e. a single stick-slip or a series of stick-slips) on flash-heating efficiency.

**Mechanical and thermocouple data** We performed laboratory earthquakes at confining pressures of 45, 90 and 180 MPa, typical pressures in the brittle crust (Table 1). Differential stress, displacement and temperature are plotted as a function of time for each stick-slips series in Supp. Fig. S1a-c. At confining pressure of 45 MPa, three stick-slips occurred along 620 μm of cumulative displacement with resulting static stress drops ranging from 12 to 20 MPa (Table 1). At confining pressure of 90 MPa, a total of 18 stick-slips occurred with an increasing release of stress between 0.5 and 23 MPa, along 2.4 mm of final slip (Table 1). At confining pressure of 180 MPa, two stick-slips occurred for a total displacement of 1.95 mm with respective stress drops around 48 MPa and 74 MPa (Table 1).

The thermocouple, located at 5.5mm away from the fault plane, detected significant co-seismic temperature increases (Supp. Fig. S1c). These are clearly observed for stress drops >5

MPa. For example, at confining pressure of 90 MPa, temperature increases only after the fifth stick-slip onwards (after the yellow arrow). Following the thermocouple inversion method described later, we focused only on stick-slips showing clear increases of temperature. The rise of temperature is around 0.075°C at the beginning of the experiment at confining pressure of 45 MPa, ranges from 0.021 to 0.086°C at 90 MPa and is around 0.4°C at 180 MPa (Fig. 1b-d and Table 1). Temperature increases together with the displacement and the static shear stress drop (Fig. 1b-d).

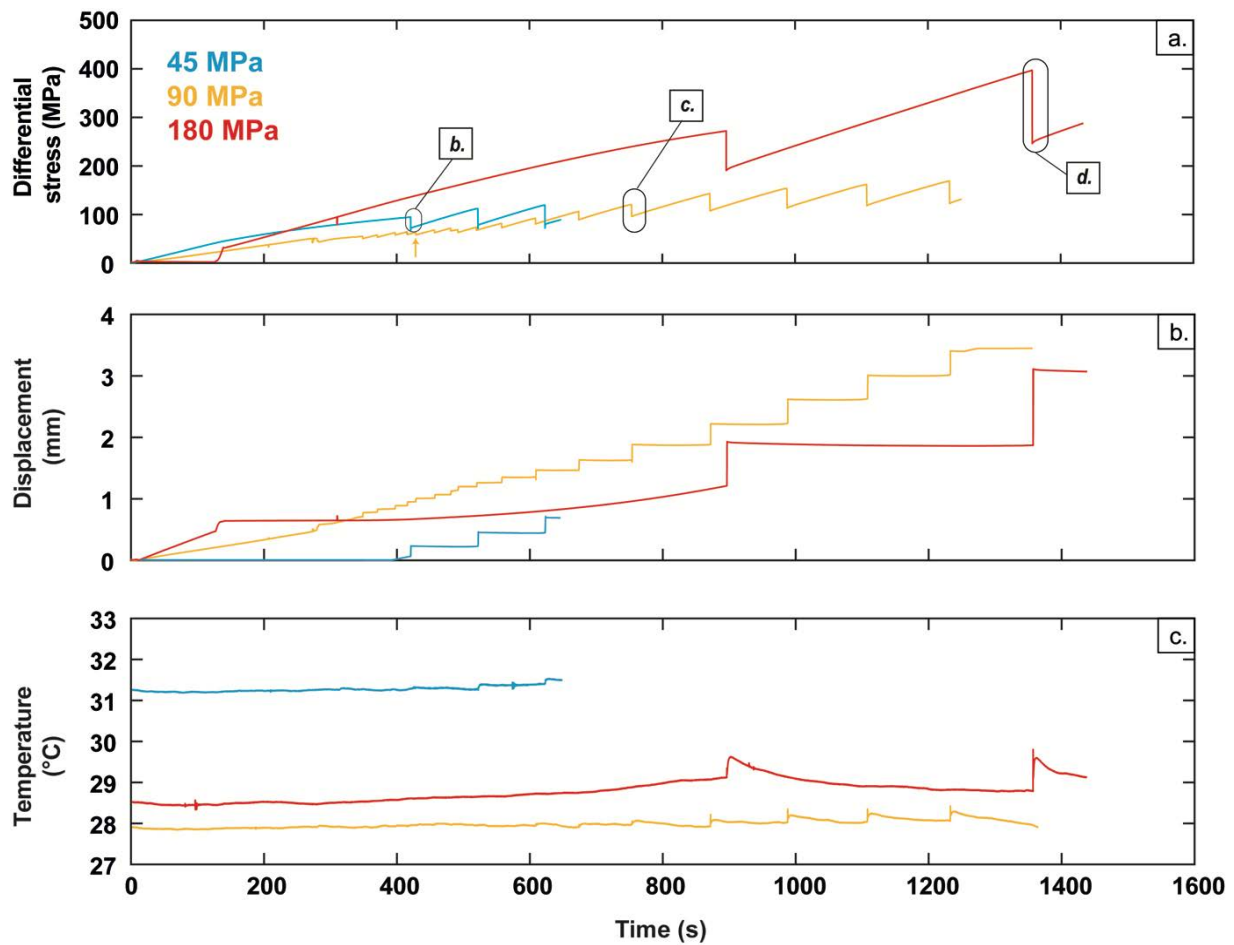

**Supp. Fig. S1. Mechanical and thermocouple data.** Differential stress (Fig. S1a), displacement (Fig. S1b) and temperature (Fig. S1c) are shown as a function of time. Boxes b, c and d in narrow bold italic are the three characteristics, for each confining pressure, described

in Fig. 1. At the onset of stick-slip, an abrupt, transient peak of temperature is observed, that is particularly large at high confining pressure. This thermal phenomenon might be explained by a local heating of the thermocouple in reply to shock waves generated by the brutal stick-slip motion[33]. As shown by the series of stick-slips, the carbon layer coated on the fault surface (thickness of a few nanometres) does not influence the displacements nor the repeatability of stick-slips compared with experimental faults[32,33] without carbon coating on the fault.

**Frictional and flash heating processes** At low confining pressures of 45 and 90 MPa, melting process is mainly related to flash heating of local asperities, as observed in previous studies[10,11,33,34]. At a larger confining pressure of 180 MPa, the size of asperities increases due to large normal stress, which promotes the complete shear heating (melting) of the fault surface.

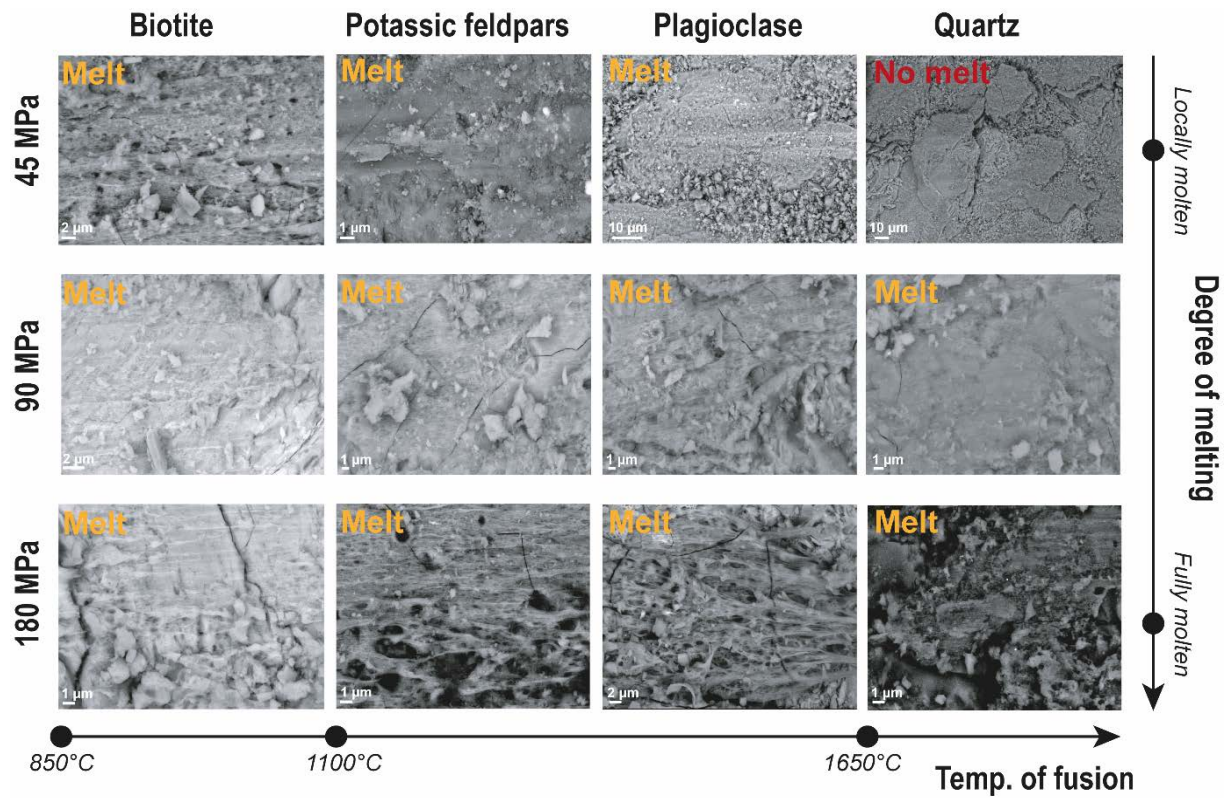

**Supp. Fig. S2. SEM micrographs of post-mortem surfaces.** Evidence of asperity melting is found on fault surfaces deformed at confining pressures of 45, 90 and 180 MPa. The degree of melting increases with the confining pressure and the surface contact area on the fault.

Fracturing processes are also observed in the sub-surface of the fault. However, clear evidence of melt injected into fractures are not found. After sliding, especially at high confining pressure, the melt layer stretches entirely along the interface (Fig. 3, Supp. Fig. S3). Progressively zooming on the melt layer, we document that it is composed of particles originating from minerals plucked along the fault interface (Supp. Fig. S3b-c), since the onset of sliding. We reason that the melt observed here corresponds to the reworking of molten gouge particles.

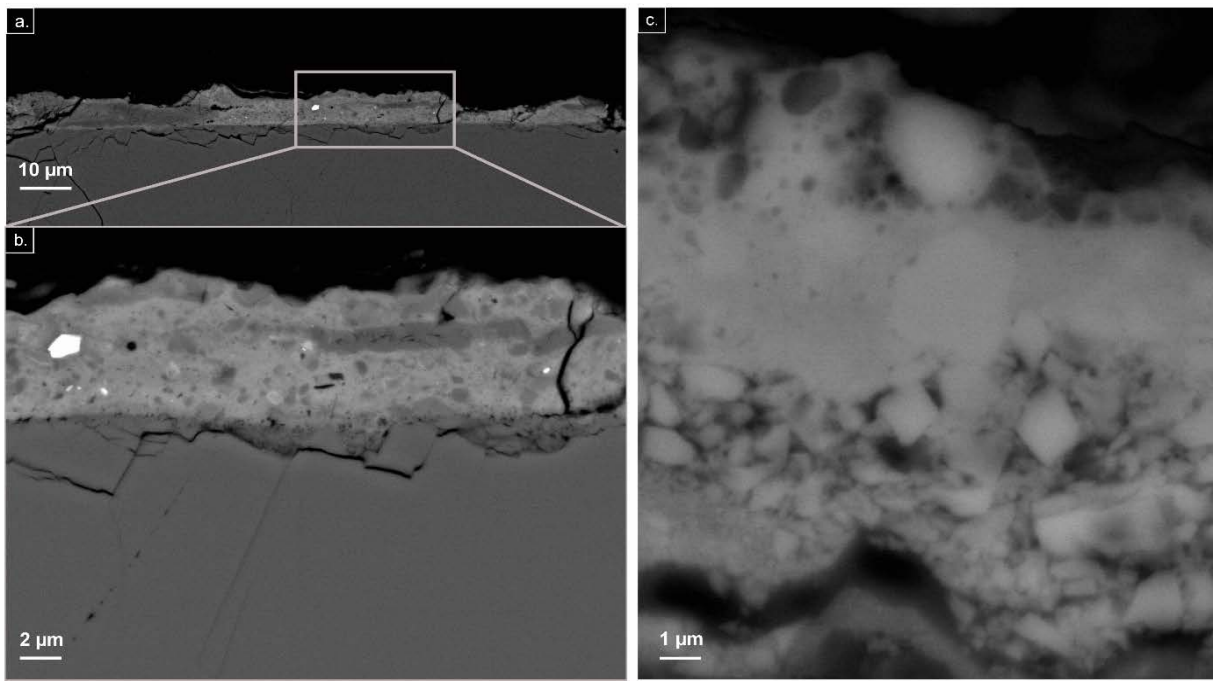

**Supp. Fig. S3. SEM micrographs of melt on the fault surface.** From (a) to (c), we zoom progressively onto the "gouge" amorphous-like material, documenting also the "chaotic"

fracturing processes that occurred during shear sliding. The melt layer is found continually along the fault interface (a and b). Melt thickness is 8–16 μm width along the fault and is composed of micro- and nano- grains (c). The experiment was performed at confining pressure of 180 MPa.

From the onset of sliding, the formation of gouge particles starts on the smooth surface, building roughness and melt network (Supp. Fig. S4a-b). This dissipates a large part of the energy by frictional heating, heating efficiency being high for low stress drops (Supp. Fig. S1 and Fig. 2b). After a few stick-slips, fault surface is intensely reworked and facilitates frictional melting, hence decreasing the heating efficiency (Fig. 2b). Furthermore, again after successive stick-slips, the gouge and melt patches are flattened and can influence the next stick-slip motion by the lubrication process (Supp. Fig. S4c-d), which thus decreases to its minimum value the heating efficiency.

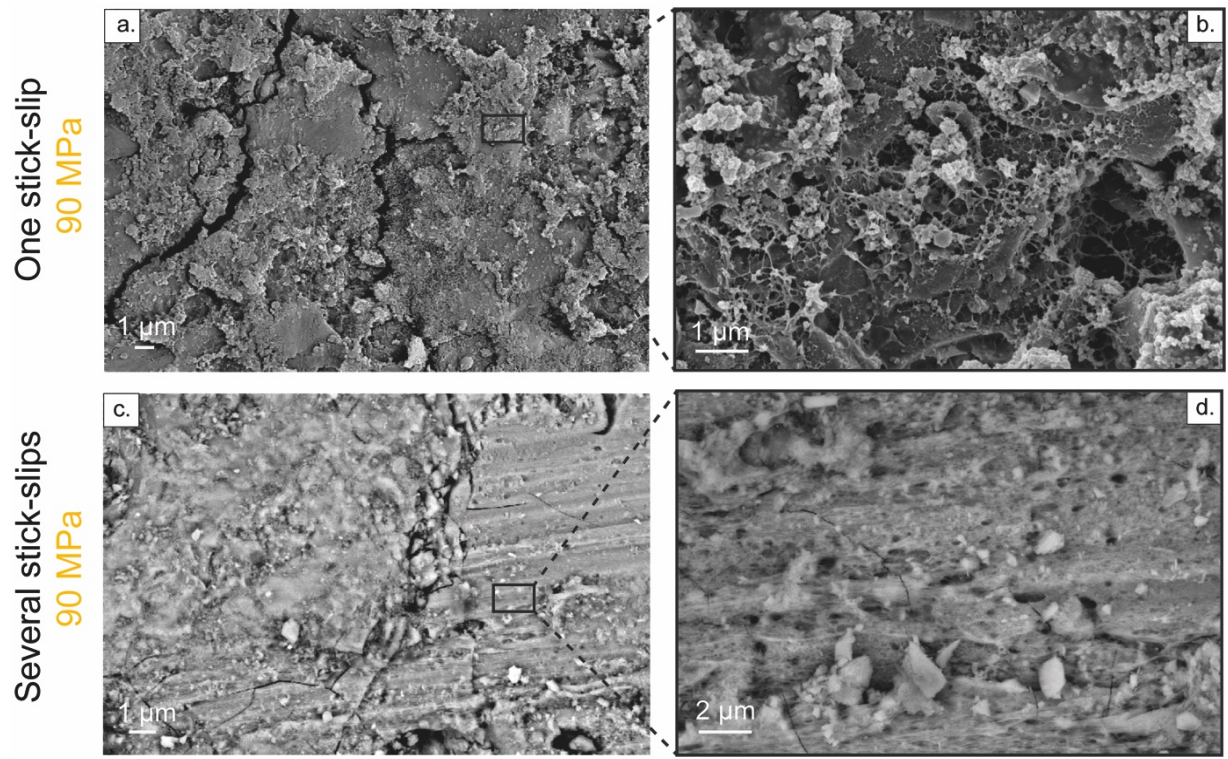

> **Supp. Fig. S4. Effects of the number of stick-slips on the fault surface.** a) and b) SEM micrographs of a fault surface after only one stick-slip at 90 MPa confining pressure. Roughness and melt network are building. c) and d) SEM micrographs conducted on the fault surface of the 90 MPa confining pressure experiment after a series of successive stick-slips.

**Calibration of the carbon technique** The estimation of temperature on the fault surface based on the coated carbon layer was calibrated using dedicated heating experiments of the amorphous carbon. A piece of amorphous carbon was heated progressively in an oven under controlled deoxygenated atmosphere (argon), for a total duration of 5 s. The carbon piece experienced carbonization and graphitization depending on the heating temperature. First, for each heating temperature, from 700°C to 1400°C, RMS measurements were done on the carbon and the $H_d/H_g$ ratio was obtained (Supp. Fig. 5a-b). Second, we studied the effect of the heating duration, from 5 s to 500 s. The values of $H_d/H_g$ ratio depend on the heating duration (Supp. Fig. 5c-d), but the difference is smaller for higher temperatures. Below 5 s of heating, none carbonization product was observed. Although in our stick-slip experiments, heating lasted approximately a few microseconds, our calibration remains robust and the temperature estimate given by RMS mapping should be considered thus as a minimum estimate, more and more accurate as temperature increases.

Nevertheless, our calibration does not consider the effect of shear strain which occurs during the stick-slip experiments, as avowed for graphite thermometer[50]. Indeed, the temperature maps (Fig. 4) show that increasing shear strain clearly influences surface temperature, because average temperature on the fault plane increases with confining pressure, as well as the molten surface area on the fault. This also suggests that temperature estimates given by the carbon method remain minimum estimates experienced during frictional sliding.

In general, 70% to 90% of the coated surface could be characterised directly by RMS mapping. When carbon spectrum was not measurable at a given location on the fault surface, we averaged the $H_d/H_g$ ratios obtained around it.

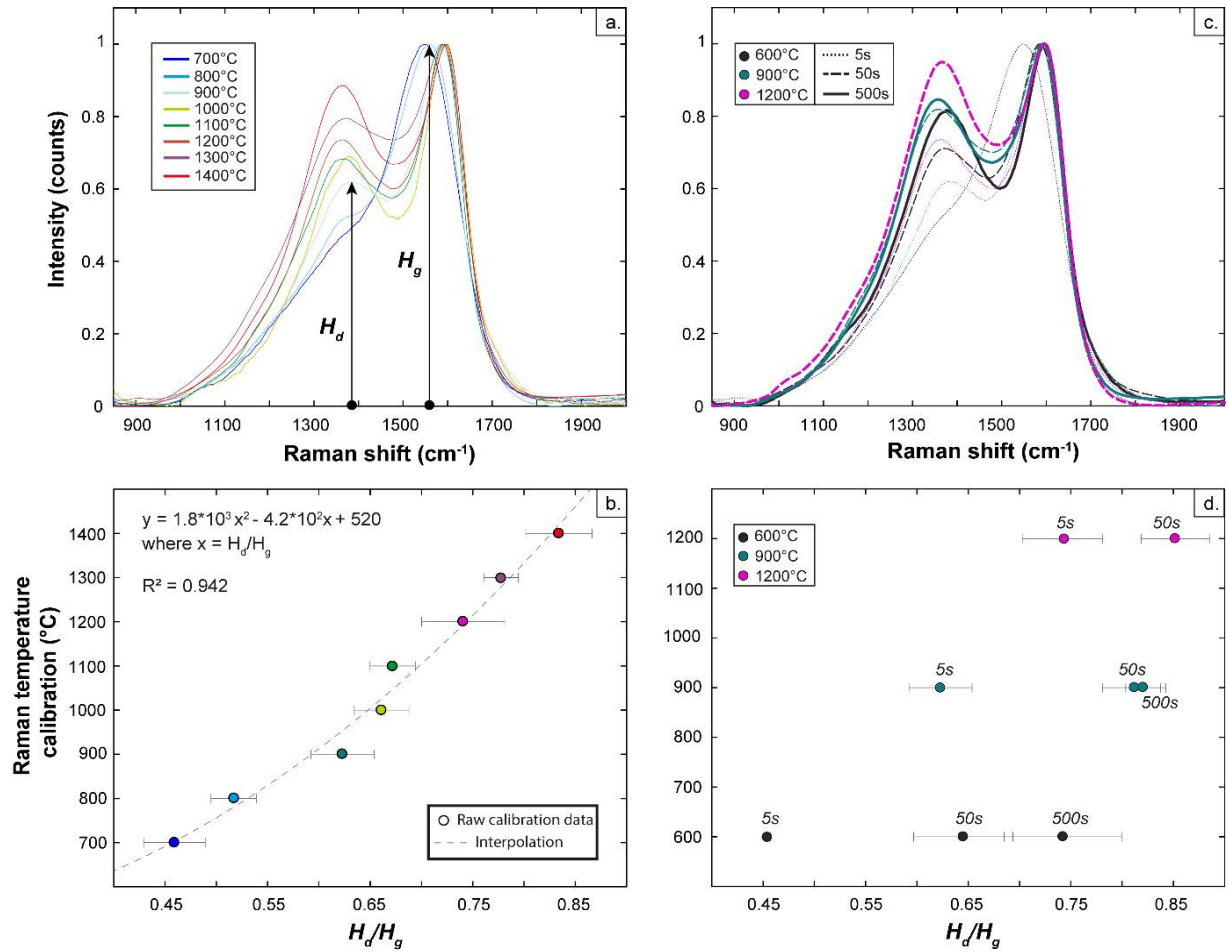

**Supp. Fig. S5. Calibration of the carbon technique for temperature estimates.** a) and b) Carbon calibration at 8 heating temperature steps from 700°C to 1400°C. When heating temperature increases, the $H_d/H_g$ ratio increases. c) and d) Influence of the heating duration on the $H_d/H_g$ ratio, for 5, 50 and 500 s of heating. The accuracy of the temperature estimate is better for higher heating temperature.

**Comparing carbon and thermocouple methods** To study the impact of a single stick-slip event compared with cumulative stick-slip events on the fault surface, we estimated the expected melted surface using the thermocouple and the carbon methods beyond the minimum temperature of fusion for biotite (850°C). In temperature maps, the temperature of fusion is given by the ratio $H_d/H_g$ on each point of the fault surface. Using the thermocouple method, the percentage of melted surface can be estimated using: %$_{melt}$=100($T_{th} - T_{amb}$)/($T_m - T_{amb}$), where $T_{th}$ is the fault temperature inverted from the thermocouple data, $T_{amb}$ is the temperature in the loading cell during the experiment (fixed to 31°C) and $T_m$ is the minimum temperature of fusion for biotite (fixed at 850°C). The melted surface inferred from the thermocouple is calculated for the largest magnitude stick-slip event and for the sum of all the cumulative stick-slip events in a given experiment.

At confining pressures of 45 and 90 MPa, the RMS %$_{melt}$ is 48% and 68%, respectively, while the thermocouple %$_{melt}$ yields only half of the RMS %$_{melt}$ (Supp. Fig. S6). At low confining pressure, a temperature of 850°C is sufficient to melt about half of the fault surface. At confining pressure of 45 MPa, cumulative sliding allows to reach 70% of melting on the surface. At 90 MPa, cumulative stick-slips melted the whole surface. At 180 MPa, a single stick-slip event is sufficient to melt the whole surface, which confirms the observations on SEM micrographs (Supp. Fig. S4). The fact that the entire fault surface melted during the experiment at 180 MPa (Fig. 3) suggests that the area of contact was almost total. This is consistent with the Hertzian contact theory. While the size of the asperities is predefined by the fault roughness and the wavelength of contacts, their size is supposed to increase linearly with increasing normal stress, reducing the wavelength of the fault roughness and increasing the area of contact.

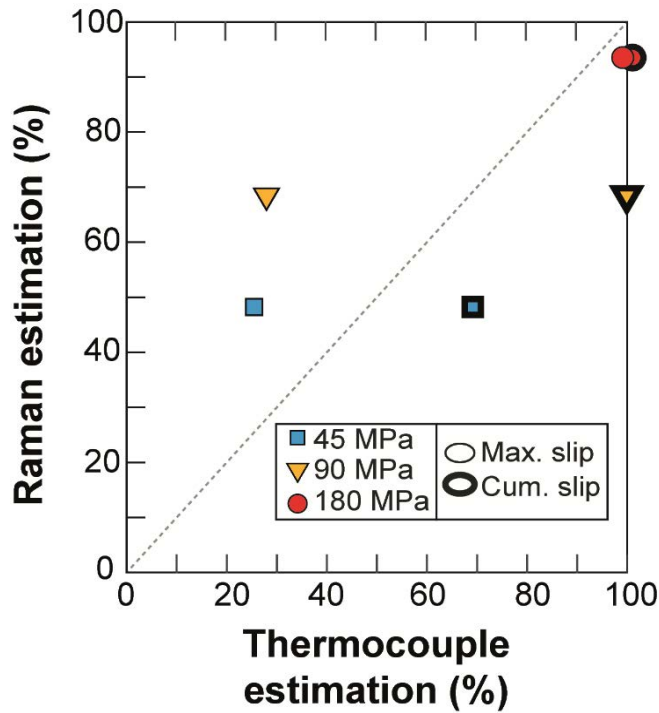

Supp. Fig. S6. Surface melted from RMS and thermocouple estimates for the minimum temperature of fusion of biotite (850°C). This melted surface is expressed as a percentage of the total surface and is calculated for the largest magnitude event and for the sum of the stick-slips in a given experiment at three confining pressures. A unique stick-slip does not allow a complete melting of the fault plane except at a confining pressure of 180 MPa. By contrast, successive stick-slips promote melting of the fault surface mainly by the reworking of asperities.

**Supplementary references (only cited in Supplementary information)**

[50] Kirilova, M., et al. Structural disorder of graphite and implications for graphite thermometry. *Journal of Geophysical Research: Solid Earth*, **9**(1)**,** 223-231 (2018).